# Band structure of magnonic crystals with defects: Brillouin spectroscopy and micromagnetic simulations

Kai Di, Vanessa Li Zhang, Meng Hau Kuok,[a)] Hock Siah Lim,[b)] Ser Choon Ng,

*Department of Physics, National University of Singapore, Singapore 117551*

Kulothungasagaran Narayanapillai, and Hyunsoo Yang[c)]

*Department of Electrical and Computer Engineering, National University of Singapore, Singapore 117576*

Using Brillouin spectroscopy, the first observation has been made of the band structures of nanostructured defect magnonic crystals. The samples are otherwise one-dimensional periodic arrays of equal-width $Ni_{80}Fe_{20}$ and cobalt nanostripes, where the defects are stripes of a different width. A dispersionless defect branch emerges within the bandgap with a frequency tunable by varying the defect stripe width, while the other branches observed are similar to those of a defect-free crystal. Micromagnetic and finite-element simulations performed unveil additional tiny bandgaps and the frequency-dependent localization of the defect mode in the vicinity of the defects.

The controlled introduction of defects in materials can radically alter their properties in desired ways. This is perhaps best exemplified by the enhanced electrical conductivity of semiconductors, as well as the color and luminescence of certain crystals, such as ruby, arising from trace amounts of chemical impurities in them. Properties of artificial crystals, in the form of periodic arrays of nanostructures, can also be controlled or enhanced by the presence of defects. Here, the defects are deviations from the artificial crystal periodicity which result in imperfect crystals. For instance, a structural or compositional defect could be a stripe element, of a different width or material, in an otherwise perfect periodic multi-component array of stripes.

Defect-induced properties of artificial photonic crystals have attracted much attention owing to their interesting physics and wide potential applications. Indeed, various photonic devices based on defects already serve as building blocks in integrated photonic circuits [1]. It is found that microcavities which are isolated defects in photonic crystals can be employed to control and manipulate the spontaneous emission of light [2,3], a functionality which allows their application in lasers and quantum optics. Other defect-induced phenomena include the all-optical control of light propagation based on enhanced optical nonlinearity in photonic-crystal nanocavities demonstrated recently [4,5]. Also, defects in photonic crystals have been shown to slow down, trap and guide light in the nanoscale [6-8].

In the case of magnonic crystals (MCs), the magnetic counterpart of photonic crystals, although extensive studies have been undertaken on defect-free MCs [9-15], relatively little research, especially experimental ones, has been done on defect MCs [16,17]. For defect-free MCs, forbidden magnonic gaps have been experimentally observed in one- and two-dimensional (1D and 2D) MCs [12-15,18]. Nikitov et al. [9] calculated the band structures and spin wave (SW) propagation properties of multilayer-type lattices with defect, while Yang et al. [19] theoretically investigated the band structure and coupling between point defects of a 2D MC. In their microwave transmission and reflection study of magnetostatic SWs in defect magnetic structures, Filimonov et al. [16] observed resonance peaks which they ascribed to defect states. The structures studied were linear 150micron-period arrays of etched grooves, on an yttrium iron garnet (YIG) film, each of which contained only either a single defect groove or crest. Chi et al. [17] showed that a line defect in a 2D MC can give rise to, as well as guide confined magnetostatic waves. In analogy to defect photonic crystals, it is expected that the incorporation of defects in MCs opens the way to new physics and new functionalities for application in data communication, processing, and magnetic storage devices. However, to date, there is no experimental observation of the band structures of nanostructured defect MCs, a crucial information for the understanding of SW propagation in these metamaterials.

Here, we report on the first observation of the band structures of nanostructured MCs with periodic defects. Brillouin light scattering (BLS) was used to detect SWs in 1D periodic arrays of alternating equal-width Permalloy ($Ni_{80}Fe_{20}$, abbreviated to Py) and cobalt nanostripes, with defects in the form of Py stripes of non-regular widths. The localization property of the defect modes and the effect of defects on the extended modes were also investigated. Time-domain micromagnetic and frequency-domain finite-element simulations were also undertaken on the dispersion relations, mode profiles and BLS intensities of magnons in the samples studied.

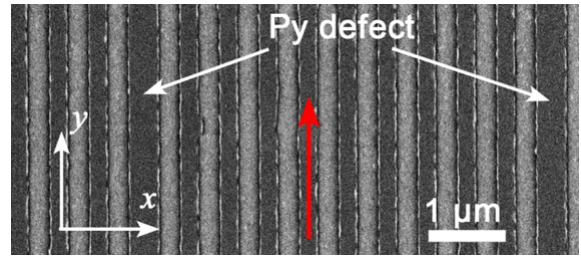

FIG. 1. SEM image of the MC with 400nm-wide Py defect stripes. Darker regions represent Py stripes while lighter regions, the Co stripes. The red arrow indicates the direction of equilibrium magnetization.

The 30nm-thick MCs, of 100×100 μm² surface area, were fabricated using electron-beam lithography, magnetron sputtering and lift-off processes [20]. The perfect MC consists of alternating 250nm-wide Py and Co stripes, with a periodicity $a$ = 500 nm. The defect MCs are otherwise perfect arrays with one Py stripe, having a width other than 250 nm, periodically located in every 10 periods. Artificial crystals containing Py defect widths of 300, 400, and 500 nm were synthesized. Figure 1 shows the scanning electron microscope (SEM) image of a defect MC, with a defect stripe width of 400 nm and a supercell period $a_d$ = 5150 nm, which will hereafter be referred to as the 400nm defect MC. Prior to the BLS measurements, the samples were saturated along the long axis of the stripes in a magnetic field $\mu_0 H = 0.4$ T, which was subsequently reduced to zero. Using the 514.5nm radiation of an argon-ion laser and a (3+3)-pass tandem Fabry-Pérot interferometer, BLS spectra were recorded in the 180º back-scattering geometry and in $ps$ polarization. With the scattering plane lying perpendicular to the long axis of the stripes, SWs propagating in the $x$-direction were probed (Fig. 1). Figure 2 presents typical Brillouin spectra of the perfect and 400nm defect MCs, recorded at various wavevectors about their respective first Brillouin zone (BZ) boundaries of $q = \pi/a$, and $q = 10\pi/a_d$. All the observed Brillouin peaks shifted under an applied magnetic field and were thus ascribed to SW modes, with the middle peak assigned to a defect mode in the defect MC (see discussion below).

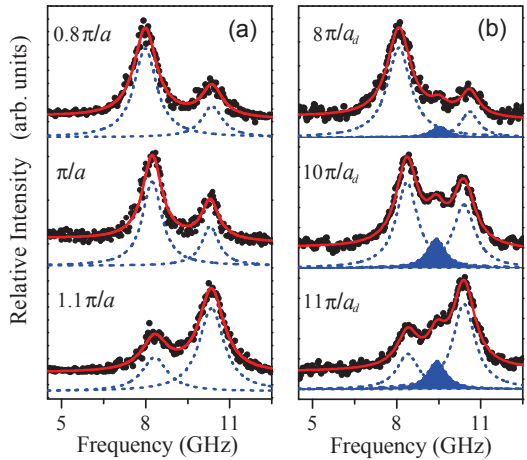

FIG. 2. Brillouin spectra of the (a) defect-free MC and (b) 400nm defect MC recorded at various wavevectors about their respective first BZ boundaries of $q = \pi/a$, and $q = 10\pi/a_d$. Measured spectra are represented by dots. All spectra were fitted with Lorentzian functions (blue dotted curves), and the resultant fitted spectra are shown as red solid curves. Defect mode peaks are shown shaded in blue.

The measured magnon dispersion relations of the defect-free MC and the 400nm defect MC are presented in Fig. 3. For the latter, besides the two prominent branches corresponding to those of the defect-free MC [12,18], an additional dispersionless branch, which is symmetrically located (≈ 9.5 GHz) about the former two, appears.

Time-domain micromagnetic simulations were performed by solving the Landau-Lifshitz-Gilbert equation [21] based on the OOMMF package [22], for which $50\mu m \times 50\mu m \times 30 nm$ MC models and $5nm \times 50\mu m \times 30nm (= \Delta x \cdot \Delta y \cdot \Delta z)$ computational cell size were used. In the calculations, the saturation magnetizations and the exchange stiffness constants of Co and Py were taken to be $M_{S,Co}$ = 1.1×10⁶ A/m, $A_{Co}$ = 2.5×10⁻¹¹ J/m, $M_{S,Py}$ = 7.3×10⁵ A/m, $A_{Py}$ = 1.1×10⁻¹¹ J/m, while the gyromagnetic ratio and the Gilbert damping coefficient were set at $\gamma$ = 195 GHz/T and $\alpha$ = 0.0001 respectively. As only SW modes in the $x$-direction were measured, modes with magnetization variation along the long axes of the stripes were neglected in the simulations, since only one cell in the $y$-direction was considered. A pulsed sinc magnetic field in the $z$-direction, $H_z \propto \sin[2\pi f(t-t_0)]/2\pi f(t-t_0)$ where $f$ = 20 GHz, was applied within one unit cell to excite spin waves. The calculated dispersion relations, obtained by a Fourier transform of the dynamic magnetization with respect to time and $x$, are presented in Fig. 3 as green lines.

Frequency-domain finite-element simulations based on the above magnetic parameters were also carried out, by solving the 2D linearized Landau-Lifshitz equation implemented in the COMSOL Multiphysics software [23], with the Bloch-Floquet boundary condition applied. A supercell comprising nine Py/Co unit cells plus a defect Py/Co cell was used in the calculation of the dispersion relations. To compare theory with experiments, the Brillouin scattering intensities of SW modes $m_{q,f}(r)$ were estimated as Fourier transforms of the out-of-plane components of their dynamic magnetization $m_z$ namely, $I \propto \left|\int e^{-iq \cdot r} m_{q,f}(r) d^3 r\right|^2$, where $q$ and $f$ are the mode wavevectors and frequencies. Finite-element calculated SW dispersion relations are presented in Fig. 3 as black circles, the sizes of which give an indication of the Brillouin intensities. Because the BZ of the supercell of the defect MC is smaller than that of the perfect MC unit cell, the theoretical band structure of the former comprises more dispersion curves. However, only modes of the branches corresponding to the two branches of the perfect MC have significant Brillouin intensities, a feature borne out by Brillouin measurements. As seen from Fig. 3, this observation is also consistent with the micromagnetic-simulated modes, whose calculated Brillouin intensity is proportional to the intensity of the shade of green representing them. It should be noted that unlike the arrays of micron-size YIG elements studied by Filimonov $et$ $al.$ [16] where only the magnetostatic dipolar interaction is taken into account, consideration of the exchange interaction for our nano-size structures is essential for good agreement with BLS experiments.

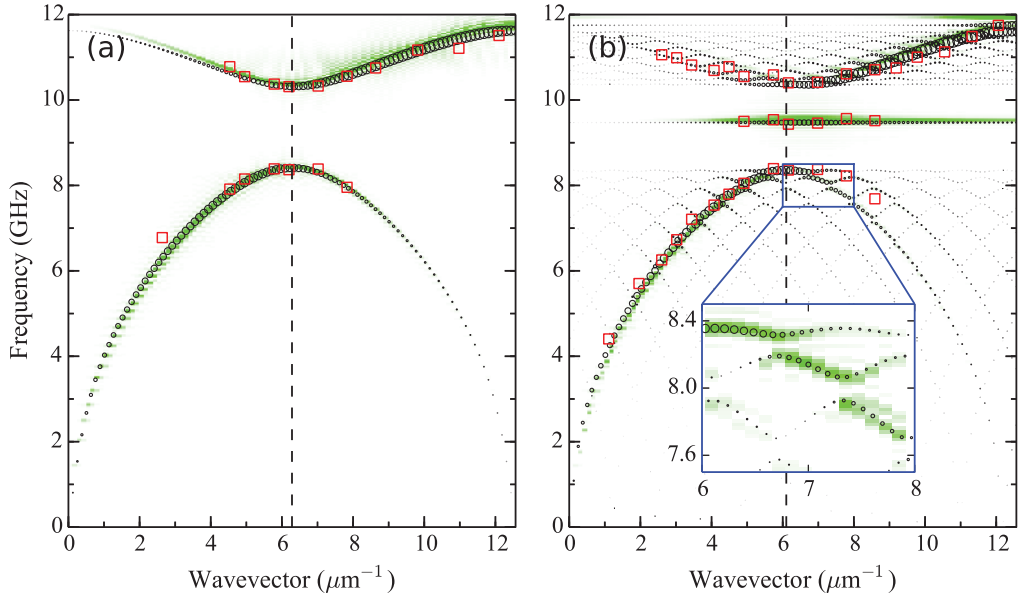

FIG. 3. Magnon band structures of (a) the perfect MC and (b) the 400nm defect MC. The dashed vertical lines indicate their respective BZ boundaries at $q = \pi/a$ and $q = 10\pi/a_d$, where $a = 500$ nm and $a_d = 5150$ nm. Measured Brillouin data are indicated by red squares, while finite-element and micromagnetic simulated data are shown as black circles and green lines, respectively. The circle size and the green shade intensity are proportional to the calculated Brillouin intensities.

Reference to Fig. 3 reveals that except for the defect mode branch, the main features of the measured and simulated band structures of the defect MC are very similar to those of the defect-free MC. This is expected as the influence of low-concentration defects on the collective magnetic excitations is weak. In the limiting case of zero defects, the defect MC should have a dispersion relation identical to that of a perfect one. It is interesting to know how the presence of low-concentration periodic defects affects the extended SW branches. Our calculations reveal that the modified periodicity of the defect MC induces additional bandgaps at the BZ boundaries corresponding to a primitive supercell, as shown in Fig. 3. These bandgaps, having narrow widths of about 0.2 GHz, were not resolved by BLS measurements because the linewidth (~ 1 GHz) of the inelastically scattered signal is much broader than these bandgaps.

The dependence of defect modes on defect Py stripe widths was also experimentally and theoretically investigated. BLS spectra of MCs with various defect stripe widths were recorded at BZ boundary wavevectors $q = 10\pi/a_d$, where $a_d$ are their respective periods. Measurements reveal that the frequency of the defect mode decreases with increasing defect stripe width, as shown in Fig. 4. In contrast, the frequencies of the bandgap edges are not very sensitive to the defect stripe width. As SWs with frequencies within the bandgap cannot propagate in the non-defect regions of an MC, the defect mode will be confined by the perfect stripes sandwiching the defect ones [9]. Because of the confinement, the narrower the defect width, the shorter would be the mode wavelength and hence, the higher its frequency. This effect is consistent with the calculations presented in Fig. 4. As the width of the defect stripe approaches the 250 nm width of a defect-free one, a transition from a localized to an extended mode occurs (see discussion below).

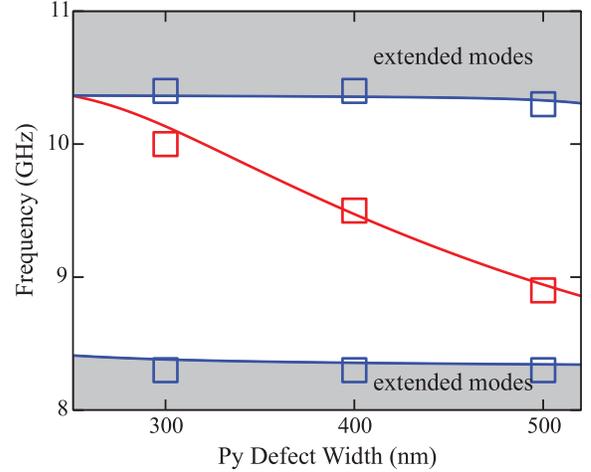

FIG. 4. Dependence of defect mode frequency on defect Py stripe width. The red squares and curve represent measured and calculated frequencies of the defect modes at the BZ boundary $q = 10\pi/a_d$, while the blue ones, the measured and calculated band edge frequencies, with the shaded regions representing the allowed magnonic bands.

The simulated defect mode profile, across one supercell, for the 400nm defect MC is displayed in Fig. 5(a). It is clear that the amplitude of the defect mode is maximal within the defect Py stripe and decays rapidly outside it, with the mode penetrating into the non-defect portions by only about four stripes. This localization behavior of the defect mode in the vicinity of the defect is consistent with earlier calculations [9,24,25]. Additionally, the defect mode profile of 300nm defect MC, shown in Fig. 5(b), indicates that the localization of a defect mode is more pronounced, the closer its frequency is to the bandgap center frequency (≈ 9.4 GHz), and it becomes more extended the closer its frequency is to those of the extended branches (see also Fig. 4). We can understand this behavior by expanding the upper extended dispersion curve $f(q)$ to lowest order near the bandgap [26]: $\Delta f \equiv f(q) - f(\pi/a) \approx c(q - \pi/a)^2$, where $c > 0$ is a constant. For a defect mode with a frequency $f(q)$ smaller than $f(\pi/a)$, $\Delta f < 0$, and thus $q \approx \pi/a \pm i\sqrt{-\Delta f/c}$. The imaginary part of $q$ causes the mode to decay exponentially in the non-defect portions of the crystal. A defect mode frequency that is closer to the gap center frequency will have a larger imaginary $q$ component, resulting in a faster decay of its dynamic magnetization and hence, a more pronounced mode localization. This behavior can be exploited to control the degree of coupling between a defect mode and other modes in the design of functional magnonic devices based on defects. It is noteworthy that the mode profiles of the extended modes themselves are also affected by the presence of defects. For the 400nm defect MC, the mode profiles of the lower and higher extended branches at the Γ point and the bandgap edges are shown in Fig. 5(c). In contrast to the defect mode, the propagating modes possess smaller amplitudes in the vicinity of the defect stripe.

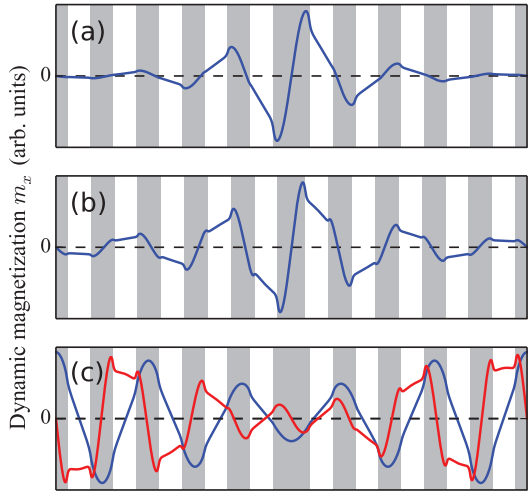

FIG. 5. Distributions of the *x*-component $m_x$ of the instantaneous dynamic magnetizations along respective supercells, at $q = 10\pi/a_d$, of (a) defect mode of the 400nm defect MC, (b) defect mode of the 300nm defect MC, and (c) the lower (blue line) and higher (red line) extended branches of the 400nm defect MC at the bandgap edges. Py and Co stripes are represented by respective grey and white bands.

Both the micromagnetic and finite-element simulations indicate that the defect mode within the bandgap can be observed by BLS only within a narrow range of wavevectors (≈ 5.5 to 8.5 μm$^{-1}$) about the first BZ boundary. This is because the distribution of the dynamic magnetization $m_z$ of the defect mode is nearly independent of wavevector. Therefore, its BLS intensity is higher for wavevectors in the vicinity of the first BZ boundary, where the Fourier transform of $m_z$ has larger components.

In summary, the defect-induced phenomena in 1D bicomponent MCs with structural defects have been investigated by BLS and numerical simulations. The presence of the Py defect stripes is principally manifested as a dispersionless branch lying within the bandgap of the experimental magnonic band structure. The two dispersive branches observed are similar to those of the defect-free MCs. Calculations also reveal that the magnon spectra of the defect crystals comprise additional branches and narrow-width Bragg bandgaps arising from the periodicity of supercell. The defect mode frequency was observed to be tunable by varying the defect stripe width, a feature that would enhance the functionality of MC-based devices, e.g. tailoring the single-frequency passband of SW filters. Just as the incorporation of defects offers novel physics and functionalities for photonic crystals, the same is expected for defects in their magnetic analogues, MCs. Our work represents the first observation of the band structures of nanostructured MCs with periodic defects and would be of use for a better understanding of their underlying physics and their applications in MC-based devices.

This project was funded by the Ministry of Education, Singapore under an Academic Research Fund Tier 1 Grant.

a)Electronic mail: phykmh@nus.edu.sg.
b)Electronic mail: phylimhs@nus.edu.sg.
c)Electronic mail: eleyang@nus.edu.sg.